\documentclass[notitlepage,twocolumn,prl,nobalancelastpage,amssymb,superscriptaddress,showpacs,floatfix]{revtex4-1}

\usepackage{epsfig}
\usepackage[colorlinks=true,citecolor=blue,linkcolor=blue]{hyperref}
\usepackage{verbatim}
\usepackage{float}
\usepackage{amsmath}
\usepackage{xcolor}
\usepackage{soul}
\usepackage{color}
\graphicspath{{fig/}}
\RequirePackage{amsmath,amsthm,amssymb,dsfont,mathrsfs}
\usepackage{amsmath }

\usepackage{graphicx}
\usepackage{yhmath}
\usepackage{mathdots}
\usepackage{MnSymbol}
\usepackage[caption=false]{subfig}
\usepackage{svg}

\begin{document}

\title{Phase Diagram of Kane-Mele Hubbard model at small doping }

\date{\today}

\author{Gaurav Kumar Gupta}
\email[]{gaurav.gupta@csun.edu\newline ggupta3@cougernet.uh.edu}
\affiliation{\mbox{Department of Physics and Astronomy, California State University, Northridge, California 91330, USA}}
\affiliation{\mbox{Texas Center for Superconductivity and Department of Physics, University of Houston, Houston, Texas 77204, USA}}
\author{D. N. Sheng}
\email[]{donna.sheng@csun.edu}
\affiliation{\mbox{Department of Physics and Astronomy, California State University, Northridge, California 91330, USA}}
\author{C. S. Ting}
\email[]{cting2@central.uh.edu}
\affiliation{\mbox{Texas Center for Superconductivity and Department of Physics, University of Houston, Houston, Texas 77204, USA}}

\begin{abstract}
Recent experiment on MoTe$_2$-WTe$_2$ twisted bi-layer demonstrated  {physics of Kane-Mele (KM) as well as Haldane models}\cite{devakul2021magic}. Although topological properties of KM model has been studied extensively, effects of interaction are still less explored beyond half filling. In this work we study the effect of Hubbard interaction in KM model at small hole doping around half filling. We use density matrix renormalization group method to characterize different phase of  cylinders with width of $L_y=3$ unit cells. We identify {a possible} superconducting (SC) phase and transition at different strength of spin-orbit coupling (SOC) and show the strong dependence of critical interaction for SC transition. We also calculate the single-particle Green's function, spin-spin and density-density correlations and compare with SC correlation to establish the dominance of SC correlations. Our result highlights  striking behaviour of SC transition and its dependence on SOC which could be useful in understanding different ways to achieve unconventional SC.	
\end{abstract}

\maketitle

{\textit{Introduction.-}}Twisted bi-layer materials have attracted a lot of attention in the recent years due to myriad of phases observed such as magnetic, correlated insulator, superconductor, zero field fractional chern insulator, among others\cite{bistritzer2011moire,andrei2020graphene,andrei2021marvels,cao2018unconventional,cao2018correlated,ledwith2020fractional, yankowitz2019tuning, sharpe2019emergent,lu2019superconductors,serlin2020intrinsic, liu2020tunable, cao2020tunable, shen2020correlated, burg2019correlated, chen2019evidence, chen2019signatures, chen2020tunable}. Other more exotic phases have been proposed theoretically such as skrymionic superconductivity  {and a superconductivity resulting from topological heavy fermion physics\cite{khalaf2021charged, song2022magic, chatterjee2020symmetry, chatterjee2022skyrmion}. 
Most of these phases arises due to the interplay of one or more pairs of flat bands, or nontrivial band topology, and interaction. These flat bands arises upon placing two or more layers of same (homo-) or different (hetero-)  two-dimensional (2D) materials with a small relative twist to form a super-cell,  which are highly tunable in terms of band width as well as band gap\cite{khalaf2021charged}. } Due to inherent nature of the large super-cell at small twist angle, chemical potential can also be tuned easily by applying external bias voltage in these materials which is otherwise impossible in most of the other materials. As these bands mostly has non-trivial band topology, recently Haldane and Kane-Mele(KM) {models}  have also been realized in twisted bi-layer transition metal dichalcogenide (tBTMD)\cite{devakul2021magic}.

Haldane model was first proposed by  Haldane in 1988 \cite{haldane1988model} where he showed that quantum Hall effect does not require external magnetic field but a topological band through breaking the time-reversal (TR) symmetry and is classified by the topological index known as Chern number. 
Later the idea was generalized by Kane-Mele(KM) \cite{kane2005z} in which two time reversal copies of the Haldane model were used together to get the first TR symmetric topological insulator and is classified by a new $\mathbb{Z}_2$ topological invariant with counter propagating edge modes under open boundary condition\cite{hasan2010colloquium, qi2011topological, bansil2016colloquium}.  

\begin{figure}[h]
	\centering
 \includesvg[width=9cm]{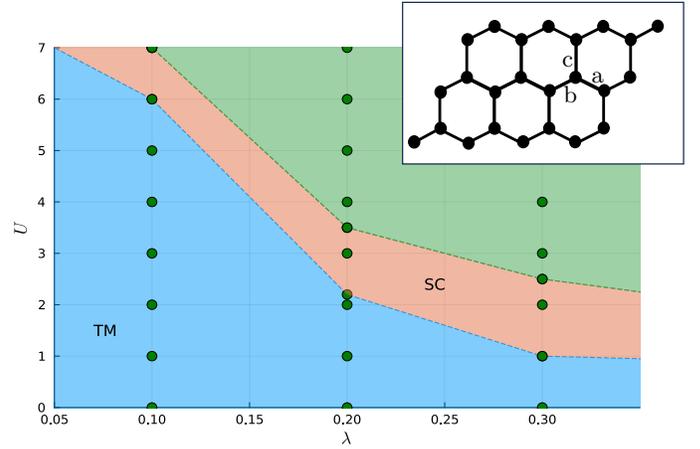}
	\caption{Quantum phase diagram of Kane-Mele Hubbard model at small hole doping ($\delta =  1/12$). The superconducting transition temperature shows a strong dependence on the spin-orbit coupling strength. For $\lambda =0.2$, the SC regime was found to be between $2.2<U<3.8$.  Inset shows the lattice used on the cylindrical geometry with open boundary condition along $x-$direction and periodic along $y-$direction for the DMRG simulation. Different bonds on which the pair correlation function was calculated has been marked as a,b, and c. Most of the calculation were performed on a system with $L_x = 24$, $L_y = 3$ with total number of  sites $N = 2\times3\times24 = 144$}.
	\label{fig:phase_diagram}
\end{figure}

Effect of interaction has also been studied in Haldane and KM models using analytical as well as numerical techniques. In Haldane model,  nearest-neighbour Coulomb interaction results in Mott insulating phase with charge density wave order\cite{mai2023topological} at strong coupling. In the spin-full Haldane Hubbard model apart from variety of  magnetically ordered phase, exotic phase which corresponds to $\mathbb{Z}_2$  fractionalized Chern insulator was proposed \cite{grushin2015characterization}.  In KM model, Hubbard interaction at   half filling  reveals an interesting phase diagram demonstrating the presence of topological band insulator, antiferromagnetic Mott insulator and a much debated quantum spin liquid phase\cite{hohenadler2012quantum, hohenadler2014phase}. Most of these studies focus on the half filling while hole or electron doped case has almost not been studied.

Inspired by the recent experiments on tBTMD which demonstrate  Kane-Mele (KM) model as well as Haldane model, and the lack of theoretical understanding of interaction away from half filling, in this work, we study the effect of interaction in KM model at small hole doping around the half filling using Density matrix renormalization group (DMRG) method\cite{white1992density}. We start with the well known KM model with Hubbard interaction on a three-leg honeycomb ladder on cylinder. We find that the critical interaction needed for  {a transition to a possible SC phase changes drastically with the change in spin-orbit coupling (SOC). We present a phase diagram  as shown in Fig. \ref{fig:phase_diagram}. For SOC strength of $\lambda = 0.2$ we find strong evidence of superconductivity for Hubbard interaction  $2.2<U<3.8$. At different SOC strengths at $0.1$ and $0.3$, we  confirm that superconductivity transition depends strongly on the SOC strength, with   critical interaction $U_c \approx 6$, and $\approx1.0$, respectively.
}

{\textit{Model and Method.-}}
Kane-Mele Hubbard (KMH) model on honeycomb lattice is given by
\begin{eqnarray}
H_{KM} = -t\sum_{\langle i,j \rangle, \sigma}c_{i\sigma}^\dagger c_{j\sigma}^{} + i \lambda\sum_{\langle\langle i,j \rangle\rangle, \sigma} \sigma c_{i\sigma}^\dagger c_{j\sigma}^{}
\end{eqnarray}
where $\langle . \rangle$ represents the nearest neighbour and $\langle\langle . \rangle\rangle$ represents the next nearest neighbour terms. $c_{i\sigma}$ ($c_{i\sigma}^\dagger$) are the  fermionic annihilation (creation) operators at site $i$ with spin $\sigma = \uparrow/\downarrow (\pm)$,  $t$ is the nearest neighbour hopping amplitude, $\lambda$ is the SOC strength. $H$ is invariant Under the TR operation.
The interacting Kane-Mele model is given by
$\mathcal{H} = H_{KM} + U\sum_{i}n_{i\uparrow}n_{i\downarrow}$,
where $U$ is the strength of the Hubbard interaction and $n_{i\sigma} = c^\dag_{i\sigma} c^{}_{i\sigma}$ are the number operator at site $i$ with spin $\sigma$. In the present study, we choose $t=1.0$ without any loss of generality and focus at the hole doping $\delta = 1/12$ near half filling.

We perform large scale DMRG simulation on honeycomb lattice with $U(1)\times U(1)$ symmetry on a cylinder with open boundary condition along $x-$direction and periodic boundary condition along $y-$direction and keep the bond dimension upto $m = 16000$ which ensures the accuracy with
a truncation error around $1.0\times 10^{-6}$. See Supplementary Materials (SM) for more details. Lattice structure used for the DMRG calculations is shown in the inset of Fig. \ref{fig:phase_diagram}. In most of the calculations, lattice dimension was fixed to $L_x = 24$ and $L_y = 3$ with the total of $N = 144$ spin-full lattice sites. 

\begin{figure}[h]
	\centering
        \includesvg[width=9cm]{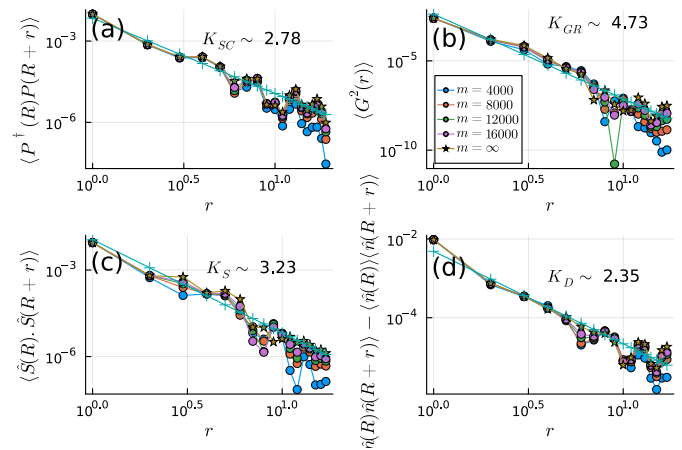}
	\caption{Quantum correlations at $U=2.5$, $\lambda = 0.2$, and $\delta = 1/12$ for bond dimensions (m) ranging from 4000 to 16000 as well as scaled bond dimension to infinity as well as fit with a power law decay to find the exponent. (a) The superconducting correlation (plotted on log-log scale) upon scaling, shows power law decay with the critical exponent $K_{SC}\sim 2.78$ (b) Single particle Green's function square ($G^2(r)$) also show power law decay with critical exponent $K_{GR}\sim 4.73$, faster decay of $G^2(r)$ indicates strong SC. (c) Spin-spin correlation function also shows faster decay with exponent $K_S\sim 3.23$ and  (d) Density-density correlation also shows decay with exponent $K_D\sim 2.35$. }
	\label{fig:corr_function_1}
\end{figure}

\begin{figure}[h]
	\centering		
        \includesvg[width=8cm]{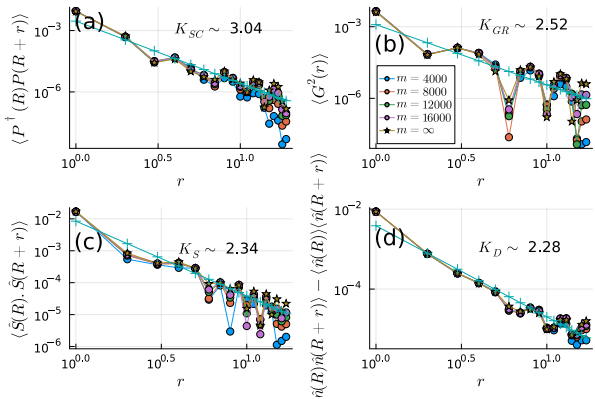}
	\caption{Quantum correlations at $U=4.2$, $\lambda = 0.2$, and $\delta = 1/12$ for bond dimensions (m) ranging from 4000 to 16000 as well as scaled bond dimension to infinity along with a fit with power law decay to find the critical exponent. (a) The superconducting correlation shows power law decay with exponent $K_{SC}\sim 3.04$.(b) Single particle Green's function square also shows power law decay with exponent $K_{GR}\sim 2.52$ which is slower than the SC correlation decy indicating the major contribution of $\langle G(r)\rangle ^2$ in $\langle P^\dagger (R) P(R + r) \rangle$ hence SC cannot be inferred in this case. (c) Spin-spin correlation function also shows power-law decay with exponent $K_S\sim 2.34$. (d) Density-density correlation function  shows power law decay with exponent $K_D\sim 2.28$.}
	\label{fig:corr_function_2}
\end{figure}

{\textit{Quantum Phase Diagram.-}} Our results are summarized in the quantum phase diagram Fig. \ref{fig:phase_diagram} as a function of interaction strength $U$ and spin-orbit coupling strength $\lambda$. For a 3-leg cylinder on honeycomb lattice with $\delta = 1/12$, we identify two phases. For $\lambda=0.2$, $U<2.2$, there is a metallic phase with partially filled $\mathbb{Z}_2$ topological band. For $2.2<U<3.5$ we show strong evidence of superconductivity. In this regime we found that the square of the single-particle Green's function is significantly smaller than pair correlation function (see Fig. \ref{fig:corr_function_1}(a-b)) indicating that pair-correlation has dominant contribution from the SC part compared to the single-particle contribution. Outside this regime, calculation show similar strength of pair-correlation and single-particle Green's function which renders the result inconclusive for SC (See Fig. \ref{fig:corr_function_2}(a-b)).

We also calculated spin-spin and density-density correlation for all the parameter regime and found no evidence of the onset of magnetic and/or charge density order. In all the cases, spin-spin and density-density correlation decays faster than pairing-correlation as shown in Fig. \ref{fig:corr_function_1}(c-d) and Fig. \ref{fig:corr_function_2} (c-d). 
 {One of our main result is
that 
the critical interaction ($U_c$) for the possible SC transition has strong dependence on $\lambda$ as shown in Fig. \ref{fig:phase_diagram}. As $\lambda$ increased, $U_c$ decreases significantly. }

{\textit{Correlation functions.-}} We define the spin-singlet pairing correlation as  $P_{\alpha,\beta}(r) = \langle \Delta_\alpha^\dagger(R) \Delta_\beta(R+r)\rangle $ where pairing term is defined as  $\Delta_\alpha(r) = (c_{r\uparrow}c_{r+\alpha\downarrow} - c_{r\downarrow}c_{r+\alpha\uparrow})/\sqrt{2}$ and $\alpha,\beta = a,b,c$ are the three different bonds on the honeycomb lattice as shown in the inset of Fig. \ref{fig:phase_diagram}. We examine the  { spin-singlet  SC correlation by looking at all the possible combination (i.e, a-a, b-b, and c-c), which is always larger than the triplet ones}. On the other hand, single particle Green's function is defined as $G(r;r^\prime) = \langle \sum_\sigma c^\dagger_{r\sigma} c^{}_{r^\prime\sigma} \rangle$. It is important to look at the single-particle Green's function along with the pairing correlation function as the pairing-correlation get a contribution from single-particle term upon applying the Wick's theorem. Along with pairing correlation and single-particle Green's function, we also calculate spin-spin correlation as well as density-density correlation which is given by $F(r) = \langle S_R. S_{R+r} \rangle$ and $D(r) = \langle n_R n_{R+r}\rangle - \langle n_R\rangle \langle n_{R+r}\rangle$ respectively. Since the correlation function in DMRG usually decays exponentially due to the finite bond dimension and inherent quasi-one-dimensional systems we consider,   we carefully perform the bond-dimension scaling to $m\rightarrow \infty$ to revel the true nature of the correlation functions.

In Fig. \ref{fig:corr_function_1}, we compare different correlation functions for $U=2.5$, $\lambda = 0.2$, and doping $\delta = 1/12$. The paring correlation grows with the bond dimension and upon bond dimension scaling to $m\rightarrow\infty$ follows a power-law decay given by $P_{aa}(r)\sim r^{-K_{SC}}$ with $K_{SC}\simeq 2.78$ as shown in Fig \ref{fig:corr_function_1} (a). At the same time, single-particle Green's function shows a significantly smaller value as well as faster decay ($K_{GR}\sim 4.73$) confirming the dominant contribution of SC correlation (Fig. \ref{fig:corr_function_1} (b)). Spin-spin correlation (Fig. \ref{fig:corr_function_1}(c) ) and density-density correlation (Fig. \ref{fig:corr_function_1} (d)) also show power-law decay faster than SC correlation and upon scaling to $m\rightarrow\infty$, we found the exponent around $3.23$ and $2.35$ respectively.

Moving on, in Fig. \ref{fig:corr_function_2} we show different correlations similar to Fig. \ref{fig:corr_function_1} for $U = 4.2$, $\lambda = 0.2$, and $\delta = 1/12$. Upon bond dimension scaling, we find  that the SC correlation still decays like a power-law with the exponent $K_{SC} \simeq 3.04$ (Fig. \ref{fig:corr_function_2} (a)) but the square of the single-particle Green's function is comparable with the SC-correlation and decays slower compared with SC-correlation ($K_{GR}\sim 2.52$). This renders the calculation inconclusive as we cannot affirmatively claim the existence of SC. We also find that spin-spin and density-density correlation decays like power law with critical exponent around $2.34$ and $2.28$ respectively. We find similar results for $U<2.2$ as discussed in the SM.

\begin{figure}[h]
	\centering
        \includesvg[width=8cm]{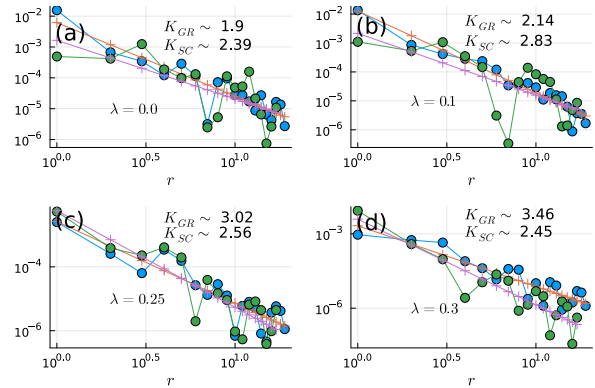}

	\caption{Quantum correlations at $U=3.0$, $\delta = 1/12$ and for different $\lambda$ between  $0.0$ and  $0.3$ with the scaled bond dimensions (m) to infinity as well as fit with a power law decay. The superconducting correlation and the square of single-particle Green's function for (a) $\lambda =0.0$, (b) $\lambda =0.1$, (c) $\lambda =0.25$, (d) $\lambda =0.3$. Different plot shows the evolution of SC correlation as the SOC strength changes. We see that for $\lambda = 0.3$ and $\lambda = 0.25$, $G^2(r)$ decays faster than  $P_{aa}(r)$ indicating strong SC while for other values of $\lambda$ ($0.0$ and $o.1$) , $G^2(r)$ decays slower than $P_{aa}(r)$, indicating the absence of SC.}
	\label{fig:sc_corr_lambda}
\end{figure}

Lastly, in Fig. \ref{fig:sc_corr_lambda} we show the strong dependence of SC transition on SOC. we found that that upon decreasing $\lambda$ to $0.1$, the SC transition appears around $U = 6.0$ which is in good agreement with the previous results of finding superconducting phase around $U\simeq 8.0 $ for pure honeycomb Hubbard model ($\lambda=0.0$)\cite{peng2023superconductivity}. On the other hand, upon increasing the SOC strength ($\lambda = 0.3$), we found that the critical interaction for SC transition decreases to $U\simeq 1.0$, which is much smaller that the $U_c \simeq 8.0$ of the original honeycomb Hubbard model.


{\textit{Summary and discussion.-}} Using large scale DMRG simulation we explore the quantum phase diagram of lightly hole doped KMH model ($\delta=1/12$). We identify SC transition and critical interaction for different values of spin-orbit coupling. For $\lambda = 0.2$, we found strong evidence of SC for the Hubbard interaction in range $2.2<U<3.8$ where the SC correlation is much more stronger than single-particle Green's function and the critical exponent is smaller than spin-spin correlation and density-density correlation. Outside this window, we found that the single-particle Green's function is comparable to the SC correlation function and hence the results are inconclusive.   Most interestingly, we found the strong dependence of SC windows in the phase diagram on the spin-orbit coupling strength. For $\lambda = 0.1$, the critical interaction for SC transition is $U_c \approx 6.0$ which is consistent with observed SC phase at $U\approx 8.0$ for pure honeycomb Hubbard model ($\lambda =0$)\cite{peng2023superconductivity}.  The critical interaction decreases significantly as we increase the SOC strength. 
We also confirm the strong dependence of SC transition on SOC strength with the help of RPA calculations.

In the present study, we studied small hole doping ($\delta = 1/12$) on KMH model, it will be interesting to study the effect of interaction on generalized KM model along with different doping. 
These results are relevant for experimental tBTMD systems, which host generalized KMH model\cite{devakul2021magic}. It will also be interesting to study the Haldane model with interaction as it is also experimentally achievable\cite{zhao2024realization}. Strong dependence of SC on SOC is also supported by RPA calculation, although we still lack more reliable theoretical knowledge for such strongly
interacting systems.  A possible future direction would be  using a better theoretical techniques such as intermediate coupling models to understand the underlying physics.  As our results are intriguing and KMH model has been realized in tBTMD, our result open new ways to find unconventional superconductivity in the field.

\begin{acknowledgments}
\section{Acknowledgments}
D.N.S. was supported by the U.S. Department of Energy, Office of Basic Energy Sciences under Grant No. DE-FG02-06ER46305. Research at University of Houston is supported by Texas Center for Superconductivity, University of Houston. The numerical calculations were performed at the Center of Advanced Computing and Data Systems at the University of Houston and University of California Northridge.
\end{acknowledgments}

{note added:
All the data and simulation code are available from the corresponding author upon reasonable request. }

\bibliography{bib.bib}

\begin{thebibliography}{34}%
\makeatletter
\providecommand \@ifxundefined [1]{%
 \@ifx{#1\undefined}
}%
\providecommand \@ifnum [1]{%
 \ifnum #1\expandafter \@firstoftwo
 \else \expandafter \@secondoftwo
 \fi
}%
\providecommand \@ifx [1]{%
 \ifx #1\expandafter \@firstoftwo
 \else \expandafter \@secondoftwo
 \fi
}%
\providecommand \natexlab [1]{#1}%
\providecommand \enquote  [1]{``#1''}%
\providecommand \bibnamefont  [1]{#1}%
\providecommand \bibfnamefont [1]{#1}%
\providecommand \citenamefont [1]{#1}%
\providecommand \href@noop [0]{\@secondoftwo}%
\providecommand \href [0]{\begingroup \@sanitize@url \@href}%
\providecommand \@href[1]{\@@startlink{#1}\@@href}%
\providecommand \@@href[1]{\endgroup#1\@@endlink}%
\providecommand \@sanitize@url [0]{\catcode `\\12\catcode `\$12\catcode
  `\&12\catcode `\#12\catcode `\^12\catcode `\_12\catcode `\%12\relax}%
\providecommand \@@startlink[1]{}%
\providecommand \@@endlink[0]{}%
\providecommand \url  [0]{\begingroup\@sanitize@url \@url }%
\providecommand \@url [1]{\endgroup\@href {#1}{\urlprefix }}%
\providecommand \urlprefix  [0]{URL }%
\providecommand \Eprint [0]{\href }%
\providecommand \doibase [0]{http://dx.doi.org/}%
\providecommand \selectlanguage [0]{\@gobble}%
\providecommand \bibinfo  [0]{\@secondoftwo}%
\providecommand \bibfield  [0]{\@secondoftwo}%
\providecommand \translation [1]{[#1]}%
\providecommand \BibitemOpen [0]{}%
\providecommand \bibitemStop [0]{}%
\providecommand \bibitemNoStop [0]{.\EOS\space}%
\providecommand \EOS [0]{\spacefactor3000\relax}%
\providecommand \BibitemShut  [1]{\csname bibitem#1\endcsname}%
\let\auto@bib@innerbib\@empty
\bibitem [{\citenamefont {Devakul}\ \emph {et~al.}(2021)\citenamefont
  {Devakul}, \citenamefont {Cr{\'e}pel}, \citenamefont {Zhang},\ and\
  \citenamefont {Fu}}]{devakul2021magic}%
  \BibitemOpen
  \bibfield  {author} {\bibinfo {author} {\bibfnamefont {T.}~\bibnamefont
  {Devakul}}, \bibinfo {author} {\bibfnamefont {V.}~\bibnamefont {Cr{\'e}pel}},
  \bibinfo {author} {\bibfnamefont {Y.}~\bibnamefont {Zhang}}, \ and\ \bibinfo
  {author} {\bibfnamefont {L.}~\bibnamefont {Fu}},\ }\href@noop {} {\bibfield
  {journal} {\bibinfo  {journal} {Nature communications}\ }\textbf {\bibinfo
  {volume} {12}},\ \bibinfo {pages} {6730} (\bibinfo {year}
  {2021})}\BibitemShut {NoStop}%
\bibitem [{\citenamefont {Bistritzer}\ and\ \citenamefont
  {MacDonald}(2011)}]{bistritzer2011moire}%
  \BibitemOpen
  \bibfield  {author} {\bibinfo {author} {\bibfnamefont {R.}~\bibnamefont
  {Bistritzer}}\ and\ \bibinfo {author} {\bibfnamefont {A.~H.}\ \bibnamefont
  {MacDonald}},\ }\href@noop {} {\bibfield  {journal} {\bibinfo  {journal}
  {Proceedings of the National Academy of Sciences}\ }\textbf {\bibinfo
  {volume} {108}},\ \bibinfo {pages} {12233} (\bibinfo {year}
  {2011})}\BibitemShut {NoStop}%
\bibitem [{\citenamefont {Andrei}\ and\ \citenamefont
  {MacDonald}(2020)}]{andrei2020graphene}%
  \BibitemOpen
  \bibfield  {author} {\bibinfo {author} {\bibfnamefont {E.~Y.}\ \bibnamefont
  {Andrei}}\ and\ \bibinfo {author} {\bibfnamefont {A.~H.}\ \bibnamefont
  {MacDonald}},\ }\href@noop {} {\bibfield  {journal} {\bibinfo  {journal}
  {Nature materials}\ }\textbf {\bibinfo {volume} {19}},\ \bibinfo {pages}
  {1265} (\bibinfo {year} {2020})}\BibitemShut {NoStop}%
\bibitem [{\citenamefont {Andrei}\ \emph {et~al.}(2021)\citenamefont {Andrei},
  \citenamefont {Efetov}, \citenamefont {Jarillo-Herrero}, \citenamefont
  {MacDonald}, \citenamefont {Mak}, \citenamefont {Senthil}, \citenamefont
  {Tutuc}, \citenamefont {Yazdani},\ and\ \citenamefont
  {Young}}]{andrei2021marvels}%
  \BibitemOpen
  \bibfield  {author} {\bibinfo {author} {\bibfnamefont {E.~Y.}\ \bibnamefont
  {Andrei}}, \bibinfo {author} {\bibfnamefont {D.~K.}\ \bibnamefont {Efetov}},
  \bibinfo {author} {\bibfnamefont {P.}~\bibnamefont {Jarillo-Herrero}},
  \bibinfo {author} {\bibfnamefont {A.~H.}\ \bibnamefont {MacDonald}}, \bibinfo
  {author} {\bibfnamefont {K.~F.}\ \bibnamefont {Mak}}, \bibinfo {author}
  {\bibfnamefont {T.}~\bibnamefont {Senthil}}, \bibinfo {author} {\bibfnamefont
  {E.}~\bibnamefont {Tutuc}}, \bibinfo {author} {\bibfnamefont
  {A.}~\bibnamefont {Yazdani}}, \ and\ \bibinfo {author} {\bibfnamefont
  {A.~F.}\ \bibnamefont {Young}},\ }\href@noop {} {\bibfield  {journal}
  {\bibinfo  {journal} {Nature Reviews Materials}\ }\textbf {\bibinfo {volume}
  {6}},\ \bibinfo {pages} {201} (\bibinfo {year} {2021})}\BibitemShut {NoStop}%
\bibitem [{\citenamefont {Cao}\ \emph {et~al.}(2018{\natexlab{a}})\citenamefont
  {Cao}, \citenamefont {Fatemi}, \citenamefont {Fang}, \citenamefont
  {Watanabe}, \citenamefont {Taniguchi}, \citenamefont {Kaxiras},\ and\
  \citenamefont {Jarillo-Herrero}}]{cao2018unconventional}%
  \BibitemOpen
  \bibfield  {author} {\bibinfo {author} {\bibfnamefont {Y.}~\bibnamefont
  {Cao}}, \bibinfo {author} {\bibfnamefont {V.}~\bibnamefont {Fatemi}},
  \bibinfo {author} {\bibfnamefont {S.}~\bibnamefont {Fang}}, \bibinfo {author}
  {\bibfnamefont {K.}~\bibnamefont {Watanabe}}, \bibinfo {author}
  {\bibfnamefont {T.}~\bibnamefont {Taniguchi}}, \bibinfo {author}
  {\bibfnamefont {E.}~\bibnamefont {Kaxiras}}, \ and\ \bibinfo {author}
  {\bibfnamefont {P.}~\bibnamefont {Jarillo-Herrero}},\ }\href@noop {}
  {\bibfield  {journal} {\bibinfo  {journal} {Nature}\ }\textbf {\bibinfo
  {volume} {556}},\ \bibinfo {pages} {43} (\bibinfo {year}
  {2018}{\natexlab{a}})}\BibitemShut {NoStop}%
\bibitem [{\citenamefont {Cao}\ \emph {et~al.}(2018{\natexlab{b}})\citenamefont
  {Cao}, \citenamefont {Fatemi}, \citenamefont {Demir}, \citenamefont {Fang},
  \citenamefont {Tomarken}, \citenamefont {Luo}, \citenamefont
  {Sanchez-Yamagishi}, \citenamefont {Watanabe}, \citenamefont {Taniguchi},
  \citenamefont {Kaxiras} \emph {et~al.}}]{cao2018correlated}%
  \BibitemOpen
  \bibfield  {author} {\bibinfo {author} {\bibfnamefont {Y.}~\bibnamefont
  {Cao}}, \bibinfo {author} {\bibfnamefont {V.}~\bibnamefont {Fatemi}},
  \bibinfo {author} {\bibfnamefont {A.}~\bibnamefont {Demir}}, \bibinfo
  {author} {\bibfnamefont {S.}~\bibnamefont {Fang}}, \bibinfo {author}
  {\bibfnamefont {S.~L.}\ \bibnamefont {Tomarken}}, \bibinfo {author}
  {\bibfnamefont {J.~Y.}\ \bibnamefont {Luo}}, \bibinfo {author} {\bibfnamefont
  {J.~D.}\ \bibnamefont {Sanchez-Yamagishi}}, \bibinfo {author} {\bibfnamefont
  {K.}~\bibnamefont {Watanabe}}, \bibinfo {author} {\bibfnamefont
  {T.}~\bibnamefont {Taniguchi}}, \bibinfo {author} {\bibfnamefont
  {E.}~\bibnamefont {Kaxiras}},  \emph {et~al.},\ }\href@noop {} {\bibfield
  {journal} {\bibinfo  {journal} {Nature}\ }\textbf {\bibinfo {volume} {556}},\
  \bibinfo {pages} {80} (\bibinfo {year} {2018}{\natexlab{b}})}\BibitemShut
  {NoStop}%
\bibitem [{\citenamefont {Ledwith}\ \emph {et~al.}(2020)\citenamefont
  {Ledwith}, \citenamefont {Tarnopolsky}, \citenamefont {Khalaf},\ and\
  \citenamefont {Vishwanath}}]{ledwith2020fractional}%
  \BibitemOpen
  \bibfield  {author} {\bibinfo {author} {\bibfnamefont {P.~J.}\ \bibnamefont
  {Ledwith}}, \bibinfo {author} {\bibfnamefont {G.}~\bibnamefont
  {Tarnopolsky}}, \bibinfo {author} {\bibfnamefont {E.}~\bibnamefont {Khalaf}},
  \ and\ \bibinfo {author} {\bibfnamefont {A.}~\bibnamefont {Vishwanath}},\
  }\href@noop {} {\bibfield  {journal} {\bibinfo  {journal} {Physical Review
  Research}\ }\textbf {\bibinfo {volume} {2}},\ \bibinfo {pages} {023237}
  (\bibinfo {year} {2020})}\BibitemShut {NoStop}%
\bibitem [{\citenamefont {Yankowitz}\ \emph {et~al.}(2019)\citenamefont
  {Yankowitz}, \citenamefont {Chen}, \citenamefont {Polshyn}, \citenamefont
  {Zhang}, \citenamefont {Watanabe}, \citenamefont {Taniguchi}, \citenamefont
  {Graf}, \citenamefont {Young},\ and\ \citenamefont
  {Dean}}]{yankowitz2019tuning}%
  \BibitemOpen
  \bibfield  {author} {\bibinfo {author} {\bibfnamefont {M.}~\bibnamefont
  {Yankowitz}}, \bibinfo {author} {\bibfnamefont {S.}~\bibnamefont {Chen}},
  \bibinfo {author} {\bibfnamefont {H.}~\bibnamefont {Polshyn}}, \bibinfo
  {author} {\bibfnamefont {Y.}~\bibnamefont {Zhang}}, \bibinfo {author}
  {\bibfnamefont {K.}~\bibnamefont {Watanabe}}, \bibinfo {author}
  {\bibfnamefont {T.}~\bibnamefont {Taniguchi}}, \bibinfo {author}
  {\bibfnamefont {D.}~\bibnamefont {Graf}}, \bibinfo {author} {\bibfnamefont
  {A.~F.}\ \bibnamefont {Young}}, \ and\ \bibinfo {author} {\bibfnamefont
  {C.~R.}\ \bibnamefont {Dean}},\ }\href@noop {} {\bibfield  {journal}
  {\bibinfo  {journal} {Science}\ }\textbf {\bibinfo {volume} {363}},\ \bibinfo
  {pages} {1059} (\bibinfo {year} {2019})}\BibitemShut {NoStop}%
\bibitem [{\citenamefont {Sharpe}\ \emph {et~al.}(2019)\citenamefont {Sharpe},
  \citenamefont {Fox}, \citenamefont {Barnard}, \citenamefont {Finney},
  \citenamefont {Watanabe}, \citenamefont {Taniguchi}, \citenamefont
  {Kastner},\ and\ \citenamefont {Goldhaber-Gordon}}]{sharpe2019emergent}%
  \BibitemOpen
  \bibfield  {author} {\bibinfo {author} {\bibfnamefont {A.~L.}\ \bibnamefont
  {Sharpe}}, \bibinfo {author} {\bibfnamefont {E.~J.}\ \bibnamefont {Fox}},
  \bibinfo {author} {\bibfnamefont {A.~W.}\ \bibnamefont {Barnard}}, \bibinfo
  {author} {\bibfnamefont {J.}~\bibnamefont {Finney}}, \bibinfo {author}
  {\bibfnamefont {K.}~\bibnamefont {Watanabe}}, \bibinfo {author}
  {\bibfnamefont {T.}~\bibnamefont {Taniguchi}}, \bibinfo {author}
  {\bibfnamefont {M.}~\bibnamefont {Kastner}}, \ and\ \bibinfo {author}
  {\bibfnamefont {D.}~\bibnamefont {Goldhaber-Gordon}},\ }\href@noop {}
  {\bibfield  {journal} {\bibinfo  {journal} {Science}\ }\textbf {\bibinfo
  {volume} {365}},\ \bibinfo {pages} {605} (\bibinfo {year}
  {2019})}\BibitemShut {NoStop}%
\bibitem [{\citenamefont {Lu}\ \emph {et~al.}(2019)\citenamefont {Lu},
  \citenamefont {Stepanov}, \citenamefont {Yang}, \citenamefont {Xie},
  \citenamefont {Aamir}, \citenamefont {Das}, \citenamefont {Urgell},
  \citenamefont {Watanabe}, \citenamefont {Taniguchi}, \citenamefont {Zhang}
  \emph {et~al.}}]{lu2019superconductors}%
  \BibitemOpen
  \bibfield  {author} {\bibinfo {author} {\bibfnamefont {X.}~\bibnamefont
  {Lu}}, \bibinfo {author} {\bibfnamefont {P.}~\bibnamefont {Stepanov}},
  \bibinfo {author} {\bibfnamefont {W.}~\bibnamefont {Yang}}, \bibinfo {author}
  {\bibfnamefont {M.}~\bibnamefont {Xie}}, \bibinfo {author} {\bibfnamefont
  {M.~A.}\ \bibnamefont {Aamir}}, \bibinfo {author} {\bibfnamefont
  {I.}~\bibnamefont {Das}}, \bibinfo {author} {\bibfnamefont {C.}~\bibnamefont
  {Urgell}}, \bibinfo {author} {\bibfnamefont {K.}~\bibnamefont {Watanabe}},
  \bibinfo {author} {\bibfnamefont {T.}~\bibnamefont {Taniguchi}}, \bibinfo
  {author} {\bibfnamefont {G.}~\bibnamefont {Zhang}},  \emph {et~al.},\
  }\href@noop {} {\bibfield  {journal} {\bibinfo  {journal} {Nature}\ }\textbf
  {\bibinfo {volume} {574}},\ \bibinfo {pages} {653} (\bibinfo {year}
  {2019})}\BibitemShut {NoStop}%
\bibitem [{\citenamefont {Serlin}\ \emph {et~al.}(2020)\citenamefont {Serlin},
  \citenamefont {Tschirhart}, \citenamefont {Polshyn}, \citenamefont {Zhang},
  \citenamefont {Zhu}, \citenamefont {Watanabe}, \citenamefont {Taniguchi},
  \citenamefont {Balents},\ and\ \citenamefont {Young}}]{serlin2020intrinsic}%
  \BibitemOpen
  \bibfield  {author} {\bibinfo {author} {\bibfnamefont {M.}~\bibnamefont
  {Serlin}}, \bibinfo {author} {\bibfnamefont {C.}~\bibnamefont {Tschirhart}},
  \bibinfo {author} {\bibfnamefont {H.}~\bibnamefont {Polshyn}}, \bibinfo
  {author} {\bibfnamefont {Y.}~\bibnamefont {Zhang}}, \bibinfo {author}
  {\bibfnamefont {J.}~\bibnamefont {Zhu}}, \bibinfo {author} {\bibfnamefont
  {K.}~\bibnamefont {Watanabe}}, \bibinfo {author} {\bibfnamefont
  {T.}~\bibnamefont {Taniguchi}}, \bibinfo {author} {\bibfnamefont
  {L.}~\bibnamefont {Balents}}, \ and\ \bibinfo {author} {\bibfnamefont
  {A.}~\bibnamefont {Young}},\ }\href@noop {} {\bibfield  {journal} {\bibinfo
  {journal} {Science}\ }\textbf {\bibinfo {volume} {367}},\ \bibinfo {pages}
  {900} (\bibinfo {year} {2020})}\BibitemShut {NoStop}%
\bibitem [{\citenamefont {Liu}\ \emph {et~al.}(2020)\citenamefont {Liu},
  \citenamefont {Hao}, \citenamefont {Khalaf}, \citenamefont {Lee},
  \citenamefont {Ronen}, \citenamefont {Yoo}, \citenamefont {Haei~Najafabadi},
  \citenamefont {Watanabe}, \citenamefont {Taniguchi}, \citenamefont
  {Vishwanath} \emph {et~al.}}]{liu2020tunable}%
  \BibitemOpen
  \bibfield  {author} {\bibinfo {author} {\bibfnamefont {X.}~\bibnamefont
  {Liu}}, \bibinfo {author} {\bibfnamefont {Z.}~\bibnamefont {Hao}}, \bibinfo
  {author} {\bibfnamefont {E.}~\bibnamefont {Khalaf}}, \bibinfo {author}
  {\bibfnamefont {J.~Y.}\ \bibnamefont {Lee}}, \bibinfo {author} {\bibfnamefont
  {Y.}~\bibnamefont {Ronen}}, \bibinfo {author} {\bibfnamefont
  {H.}~\bibnamefont {Yoo}}, \bibinfo {author} {\bibfnamefont {D.}~\bibnamefont
  {Haei~Najafabadi}}, \bibinfo {author} {\bibfnamefont {K.}~\bibnamefont
  {Watanabe}}, \bibinfo {author} {\bibfnamefont {T.}~\bibnamefont {Taniguchi}},
  \bibinfo {author} {\bibfnamefont {A.}~\bibnamefont {Vishwanath}},  \emph
  {et~al.},\ }\href@noop {} {\bibfield  {journal} {\bibinfo  {journal}
  {Nature}\ }\textbf {\bibinfo {volume} {583}},\ \bibinfo {pages} {221}
  (\bibinfo {year} {2020})}\BibitemShut {NoStop}%
\bibitem [{\citenamefont {Cao}\ \emph {et~al.}(2020)\citenamefont {Cao},
  \citenamefont {Rodan-Legrain}, \citenamefont {Rubies-Bigorda}, \citenamefont
  {Park}, \citenamefont {Watanabe}, \citenamefont {Taniguchi},\ and\
  \citenamefont {Jarillo-Herrero}}]{cao2020tunable}%
  \BibitemOpen
  \bibfield  {author} {\bibinfo {author} {\bibfnamefont {Y.}~\bibnamefont
  {Cao}}, \bibinfo {author} {\bibfnamefont {D.}~\bibnamefont {Rodan-Legrain}},
  \bibinfo {author} {\bibfnamefont {O.}~\bibnamefont {Rubies-Bigorda}},
  \bibinfo {author} {\bibfnamefont {J.~M.}\ \bibnamefont {Park}}, \bibinfo
  {author} {\bibfnamefont {K.}~\bibnamefont {Watanabe}}, \bibinfo {author}
  {\bibfnamefont {T.}~\bibnamefont {Taniguchi}}, \ and\ \bibinfo {author}
  {\bibfnamefont {P.}~\bibnamefont {Jarillo-Herrero}},\ }\href@noop {}
  {\bibfield  {journal} {\bibinfo  {journal} {Nature}\ }\textbf {\bibinfo
  {volume} {583}},\ \bibinfo {pages} {215} (\bibinfo {year}
  {2020})}\BibitemShut {NoStop}%
\bibitem [{\citenamefont {Shen}\ \emph {et~al.}(2020)\citenamefont {Shen},
  \citenamefont {Chu}, \citenamefont {Wu}, \citenamefont {Li}, \citenamefont
  {Wang}, \citenamefont {Zhao}, \citenamefont {Tang}, \citenamefont {Liu},
  \citenamefont {Tian}, \citenamefont {Watanabe} \emph
  {et~al.}}]{shen2020correlated}%
  \BibitemOpen
  \bibfield  {author} {\bibinfo {author} {\bibfnamefont {C.}~\bibnamefont
  {Shen}}, \bibinfo {author} {\bibfnamefont {Y.}~\bibnamefont {Chu}}, \bibinfo
  {author} {\bibfnamefont {Q.}~\bibnamefont {Wu}}, \bibinfo {author}
  {\bibfnamefont {N.}~\bibnamefont {Li}}, \bibinfo {author} {\bibfnamefont
  {S.}~\bibnamefont {Wang}}, \bibinfo {author} {\bibfnamefont {Y.}~\bibnamefont
  {Zhao}}, \bibinfo {author} {\bibfnamefont {J.}~\bibnamefont {Tang}}, \bibinfo
  {author} {\bibfnamefont {J.}~\bibnamefont {Liu}}, \bibinfo {author}
  {\bibfnamefont {J.}~\bibnamefont {Tian}}, \bibinfo {author} {\bibfnamefont
  {K.}~\bibnamefont {Watanabe}},  \emph {et~al.},\ }\href@noop {} {\bibfield
  {journal} {\bibinfo  {journal} {Nature Physics}\ }\textbf {\bibinfo {volume}
  {16}},\ \bibinfo {pages} {520} (\bibinfo {year} {2020})}\BibitemShut
  {NoStop}%
\bibitem [{\citenamefont {Burg}\ \emph {et~al.}(2019)\citenamefont {Burg},
  \citenamefont {Zhu}, \citenamefont {Taniguchi}, \citenamefont {Watanabe},
  \citenamefont {MacDonald},\ and\ \citenamefont {Tutuc}}]{burg2019correlated}%
  \BibitemOpen
  \bibfield  {author} {\bibinfo {author} {\bibfnamefont {G.~W.}\ \bibnamefont
  {Burg}}, \bibinfo {author} {\bibfnamefont {J.}~\bibnamefont {Zhu}}, \bibinfo
  {author} {\bibfnamefont {T.}~\bibnamefont {Taniguchi}}, \bibinfo {author}
  {\bibfnamefont {K.}~\bibnamefont {Watanabe}}, \bibinfo {author}
  {\bibfnamefont {A.~H.}\ \bibnamefont {MacDonald}}, \ and\ \bibinfo {author}
  {\bibfnamefont {E.}~\bibnamefont {Tutuc}},\ }\href@noop {} {\bibfield
  {journal} {\bibinfo  {journal} {Physical review letters}\ }\textbf {\bibinfo
  {volume} {123}},\ \bibinfo {pages} {197702} (\bibinfo {year}
  {2019})}\BibitemShut {NoStop}%
\bibitem [{\citenamefont {Chen}\ \emph
  {et~al.}(2019{\natexlab{a}})\citenamefont {Chen}, \citenamefont {Jiang},
  \citenamefont {Wu}, \citenamefont {Lyu}, \citenamefont {Li}, \citenamefont
  {Chittari}, \citenamefont {Watanabe}, \citenamefont {Taniguchi},
  \citenamefont {Shi}, \citenamefont {Jung} \emph {et~al.}}]{chen2019evidence}%
  \BibitemOpen
  \bibfield  {author} {\bibinfo {author} {\bibfnamefont {G.}~\bibnamefont
  {Chen}}, \bibinfo {author} {\bibfnamefont {L.}~\bibnamefont {Jiang}},
  \bibinfo {author} {\bibfnamefont {S.}~\bibnamefont {Wu}}, \bibinfo {author}
  {\bibfnamefont {B.}~\bibnamefont {Lyu}}, \bibinfo {author} {\bibfnamefont
  {H.}~\bibnamefont {Li}}, \bibinfo {author} {\bibfnamefont {B.~L.}\
  \bibnamefont {Chittari}}, \bibinfo {author} {\bibfnamefont {K.}~\bibnamefont
  {Watanabe}}, \bibinfo {author} {\bibfnamefont {T.}~\bibnamefont {Taniguchi}},
  \bibinfo {author} {\bibfnamefont {Z.}~\bibnamefont {Shi}}, \bibinfo {author}
  {\bibfnamefont {J.}~\bibnamefont {Jung}},  \emph {et~al.},\ }\href@noop {}
  {\bibfield  {journal} {\bibinfo  {journal} {Nature Physics}\ }\textbf
  {\bibinfo {volume} {15}},\ \bibinfo {pages} {237} (\bibinfo {year}
  {2019}{\natexlab{a}})}\BibitemShut {NoStop}%
\bibitem [{\citenamefont {Chen}\ \emph
  {et~al.}(2019{\natexlab{b}})\citenamefont {Chen}, \citenamefont {Sharpe},
  \citenamefont {Gallagher}, \citenamefont {Rosen}, \citenamefont {Fox},
  \citenamefont {Jiang}, \citenamefont {Lyu}, \citenamefont {Li}, \citenamefont
  {Watanabe}, \citenamefont {Taniguchi} \emph {et~al.}}]{chen2019signatures}%
  \BibitemOpen
  \bibfield  {author} {\bibinfo {author} {\bibfnamefont {G.}~\bibnamefont
  {Chen}}, \bibinfo {author} {\bibfnamefont {A.~L.}\ \bibnamefont {Sharpe}},
  \bibinfo {author} {\bibfnamefont {P.}~\bibnamefont {Gallagher}}, \bibinfo
  {author} {\bibfnamefont {I.~T.}\ \bibnamefont {Rosen}}, \bibinfo {author}
  {\bibfnamefont {E.~J.}\ \bibnamefont {Fox}}, \bibinfo {author} {\bibfnamefont
  {L.}~\bibnamefont {Jiang}}, \bibinfo {author} {\bibfnamefont
  {B.}~\bibnamefont {Lyu}}, \bibinfo {author} {\bibfnamefont {H.}~\bibnamefont
  {Li}}, \bibinfo {author} {\bibfnamefont {K.}~\bibnamefont {Watanabe}},
  \bibinfo {author} {\bibfnamefont {T.}~\bibnamefont {Taniguchi}},  \emph
  {et~al.},\ }\href@noop {} {\bibfield  {journal} {\bibinfo  {journal}
  {Nature}\ }\textbf {\bibinfo {volume} {572}},\ \bibinfo {pages} {215}
  (\bibinfo {year} {2019}{\natexlab{b}})}\BibitemShut {NoStop}%
\bibitem [{\citenamefont {Chen}\ \emph {et~al.}(2020)\citenamefont {Chen},
  \citenamefont {Sharpe}, \citenamefont {Fox}, \citenamefont {Zhang},
  \citenamefont {Wang}, \citenamefont {Jiang}, \citenamefont {Lyu},
  \citenamefont {Li}, \citenamefont {Watanabe}, \citenamefont {Taniguchi} \emph
  {et~al.}}]{chen2020tunable}%
  \BibitemOpen
  \bibfield  {author} {\bibinfo {author} {\bibfnamefont {G.}~\bibnamefont
  {Chen}}, \bibinfo {author} {\bibfnamefont {A.~L.}\ \bibnamefont {Sharpe}},
  \bibinfo {author} {\bibfnamefont {E.~J.}\ \bibnamefont {Fox}}, \bibinfo
  {author} {\bibfnamefont {Y.-H.}\ \bibnamefont {Zhang}}, \bibinfo {author}
  {\bibfnamefont {S.}~\bibnamefont {Wang}}, \bibinfo {author} {\bibfnamefont
  {L.}~\bibnamefont {Jiang}}, \bibinfo {author} {\bibfnamefont
  {B.}~\bibnamefont {Lyu}}, \bibinfo {author} {\bibfnamefont {H.}~\bibnamefont
  {Li}}, \bibinfo {author} {\bibfnamefont {K.}~\bibnamefont {Watanabe}},
  \bibinfo {author} {\bibfnamefont {T.}~\bibnamefont {Taniguchi}},  \emph
  {et~al.},\ }\href@noop {} {\bibfield  {journal} {\bibinfo  {journal}
  {Nature}\ }\textbf {\bibinfo {volume} {579}},\ \bibinfo {pages} {56}
  (\bibinfo {year} {2020})}\BibitemShut {NoStop}%
\bibitem [{\citenamefont {Khalaf}\ \emph {et~al.}(2021)\citenamefont {Khalaf},
  \citenamefont {Chatterjee}, \citenamefont {Bultinck}, \citenamefont
  {Zaletel},\ and\ \citenamefont {Vishwanath}}]{khalaf2021charged}%
  \BibitemOpen
  \bibfield  {author} {\bibinfo {author} {\bibfnamefont {E.}~\bibnamefont
  {Khalaf}}, \bibinfo {author} {\bibfnamefont {S.}~\bibnamefont {Chatterjee}},
  \bibinfo {author} {\bibfnamefont {N.}~\bibnamefont {Bultinck}}, \bibinfo
  {author} {\bibfnamefont {M.~P.}\ \bibnamefont {Zaletel}}, \ and\ \bibinfo
  {author} {\bibfnamefont {A.}~\bibnamefont {Vishwanath}},\ }\href@noop {}
  {\bibfield  {journal} {\bibinfo  {journal} {Science advances}\ }\textbf
  {\bibinfo {volume} {7}},\ \bibinfo {pages} {eabf5299} (\bibinfo {year}
  {2021})}\BibitemShut {NoStop}%
\bibitem [{\citenamefont {Song}\ and\ \citenamefont
  {Bernevig}(2022)}]{song2022magic}%
  \BibitemOpen
  \bibfield  {author} {\bibinfo {author} {\bibfnamefont {Z.-D.}\ \bibnamefont
  {Song}}\ and\ \bibinfo {author} {\bibfnamefont {B.~A.}\ \bibnamefont
  {Bernevig}},\ }\href@noop {} {\bibfield  {journal} {\bibinfo  {journal}
  {Physical review letters}\ }\textbf {\bibinfo {volume} {129}},\ \bibinfo
  {pages} {047601} (\bibinfo {year} {2022})}\BibitemShut {NoStop}%
\bibitem [{\citenamefont {Chatterjee}\ \emph {et~al.}(2020)\citenamefont
  {Chatterjee}, \citenamefont {Bultinck},\ and\ \citenamefont
  {Zaletel}}]{chatterjee2020symmetry}%
  \BibitemOpen
  \bibfield  {author} {\bibinfo {author} {\bibfnamefont {S.}~\bibnamefont
  {Chatterjee}}, \bibinfo {author} {\bibfnamefont {N.}~\bibnamefont
  {Bultinck}}, \ and\ \bibinfo {author} {\bibfnamefont {M.~P.}\ \bibnamefont
  {Zaletel}},\ }\href@noop {} {\bibfield  {journal} {\bibinfo  {journal}
  {Physical Review B}\ }\textbf {\bibinfo {volume} {101}},\ \bibinfo {pages}
  {165141} (\bibinfo {year} {2020})}\BibitemShut {NoStop}%
\bibitem [{\citenamefont {Chatterjee}\ \emph {et~al.}(2022)\citenamefont
  {Chatterjee}, \citenamefont {Ippoliti},\ and\ \citenamefont
  {Zaletel}}]{chatterjee2022skyrmion}%
  \BibitemOpen
  \bibfield  {author} {\bibinfo {author} {\bibfnamefont {S.}~\bibnamefont
  {Chatterjee}}, \bibinfo {author} {\bibfnamefont {M.}~\bibnamefont
  {Ippoliti}}, \ and\ \bibinfo {author} {\bibfnamefont {M.~P.}\ \bibnamefont
  {Zaletel}},\ }\href@noop {} {\bibfield  {journal} {\bibinfo  {journal}
  {Physical review B}\ }\textbf {\bibinfo {volume} {106}},\ \bibinfo {pages}
  {035421} (\bibinfo {year} {2022})}\BibitemShut {NoStop}%
\bibitem [{\citenamefont {Haldane}(1988)}]{haldane1988model}%
  \BibitemOpen
  \bibfield  {author} {\bibinfo {author} {\bibfnamefont {F.~D.~M.}\
  \bibnamefont {Haldane}},\ }\href@noop {} {\bibfield  {journal} {\bibinfo
  {journal} {Physical review letters}\ }\textbf {\bibinfo {volume} {61}},\
  \bibinfo {pages} {2015} (\bibinfo {year} {1988})}\BibitemShut {NoStop}%
\bibitem [{\citenamefont {Kane}\ and\ \citenamefont {Mele}(2005)}]{kane2005z}%
  \BibitemOpen
  \bibfield  {author} {\bibinfo {author} {\bibfnamefont {C.~L.}\ \bibnamefont
  {Kane}}\ and\ \bibinfo {author} {\bibfnamefont {E.~J.}\ \bibnamefont
  {Mele}},\ }\href@noop {} {\bibfield  {journal} {\bibinfo  {journal} {Physical
  review letters}\ }\textbf {\bibinfo {volume} {95}},\ \bibinfo {pages}
  {146802} (\bibinfo {year} {2005})}\BibitemShut {NoStop}%
\bibitem [{\citenamefont {Hasan}\ and\ \citenamefont
  {Kane}(2010)}]{hasan2010colloquium}%
  \BibitemOpen
  \bibfield  {author} {\bibinfo {author} {\bibfnamefont {M.~Z.}\ \bibnamefont
  {Hasan}}\ and\ \bibinfo {author} {\bibfnamefont {C.~L.}\ \bibnamefont
  {Kane}},\ }\href@noop {} {\bibfield  {journal} {\bibinfo  {journal} {Reviews
  of modern physics}\ }\textbf {\bibinfo {volume} {82}},\ \bibinfo {pages}
  {3045} (\bibinfo {year} {2010})}\BibitemShut {NoStop}%
\bibitem [{\citenamefont {Qi}\ and\ \citenamefont
  {Zhang}(2011)}]{qi2011topological}%
  \BibitemOpen
  \bibfield  {author} {\bibinfo {author} {\bibfnamefont {X.-L.}\ \bibnamefont
  {Qi}}\ and\ \bibinfo {author} {\bibfnamefont {S.-C.}\ \bibnamefont {Zhang}},\
  }\href@noop {} {\bibfield  {journal} {\bibinfo  {journal} {Reviews of modern
  physics}\ }\textbf {\bibinfo {volume} {83}},\ \bibinfo {pages} {1057}
  (\bibinfo {year} {2011})}\BibitemShut {NoStop}%
\bibitem [{\citenamefont {Lin}\ and\ \citenamefont
  {Das}(2016)}]{bansil2016colloquium}%
  \BibitemOpen
  \bibfield  {author} {\bibinfo {author} {\bibfnamefont {H.}~\bibnamefont
  {Lin}}\ and\ \bibinfo {author} {\bibfnamefont {T.}~\bibnamefont {Das}},\
  }\href@noop {} {\bibfield  {journal} {\bibinfo  {journal} {Reviews of Modern
  Physics}\ }\textbf {\bibinfo {volume} {88}},\ \bibinfo {pages} {021004}
  (\bibinfo {year} {2016})}\BibitemShut {NoStop}%
\bibitem [{\citenamefont {Mai}\ \emph {et~al.}(2023)\citenamefont {Mai},
  \citenamefont {Feldman},\ and\ \citenamefont
  {Phillips}}]{mai2023topological}%
  \BibitemOpen
  \bibfield  {author} {\bibinfo {author} {\bibfnamefont {P.}~\bibnamefont
  {Mai}}, \bibinfo {author} {\bibfnamefont {B.~E.}\ \bibnamefont {Feldman}}, \
  and\ \bibinfo {author} {\bibfnamefont {P.~W.}\ \bibnamefont {Phillips}},\
  }\href@noop {} {\bibfield  {journal} {\bibinfo  {journal} {Physical Review
  Research}\ }\textbf {\bibinfo {volume} {5}},\ \bibinfo {pages} {013162}
  (\bibinfo {year} {2023})}\BibitemShut {NoStop}%
\bibitem [{\citenamefont {Grushin}\ \emph {et~al.}(2015)\citenamefont
  {Grushin}, \citenamefont {Motruk}, \citenamefont {Zaletel},\ and\
  \citenamefont {Pollmann}}]{grushin2015characterization}%
  \BibitemOpen
  \bibfield  {author} {\bibinfo {author} {\bibfnamefont {A.~G.}\ \bibnamefont
  {Grushin}}, \bibinfo {author} {\bibfnamefont {J.}~\bibnamefont {Motruk}},
  \bibinfo {author} {\bibfnamefont {M.~P.}\ \bibnamefont {Zaletel}}, \ and\
  \bibinfo {author} {\bibfnamefont {F.}~\bibnamefont {Pollmann}},\ }\href@noop
  {} {\bibfield  {journal} {\bibinfo  {journal} {Physical Review B}\ }\textbf
  {\bibinfo {volume} {91}},\ \bibinfo {pages} {035136} (\bibinfo {year}
  {2015})}\BibitemShut {NoStop}%
\bibitem [{\citenamefont {Hohenadler}\ \emph {et~al.}(2012)\citenamefont
  {Hohenadler}, \citenamefont {Meng}, \citenamefont {Lang}, \citenamefont
  {Wessel}, \citenamefont {Muramatsu},\ and\ \citenamefont
  {Assaad}}]{hohenadler2012quantum}%
  \BibitemOpen
  \bibfield  {author} {\bibinfo {author} {\bibfnamefont {M.}~\bibnamefont
  {Hohenadler}}, \bibinfo {author} {\bibfnamefont {Z.}~\bibnamefont {Meng}},
  \bibinfo {author} {\bibfnamefont {T.}~\bibnamefont {Lang}}, \bibinfo {author}
  {\bibfnamefont {S.}~\bibnamefont {Wessel}}, \bibinfo {author} {\bibfnamefont
  {A.}~\bibnamefont {Muramatsu}}, \ and\ \bibinfo {author} {\bibfnamefont
  {F.}~\bibnamefont {Assaad}},\ }\href@noop {} {\bibfield  {journal} {\bibinfo
  {journal} {Physical Review B}\ }\textbf {\bibinfo {volume} {85}},\ \bibinfo
  {pages} {115132} (\bibinfo {year} {2012})}\BibitemShut {NoStop}%
\bibitem [{\citenamefont {Hohenadler}\ \emph {et~al.}(2014)\citenamefont
  {Hohenadler}, \citenamefont {Parisen~Toldin}, \citenamefont {Herbut},\ and\
  \citenamefont {Assaad}}]{hohenadler2014phase}%
  \BibitemOpen
  \bibfield  {author} {\bibinfo {author} {\bibfnamefont {M.}~\bibnamefont
  {Hohenadler}}, \bibinfo {author} {\bibfnamefont {F.}~\bibnamefont
  {Parisen~Toldin}}, \bibinfo {author} {\bibfnamefont {I.}~\bibnamefont
  {Herbut}}, \ and\ \bibinfo {author} {\bibfnamefont {F.}~\bibnamefont
  {Assaad}},\ }\href@noop {} {\bibfield  {journal} {\bibinfo  {journal}
  {Physical Review B}\ }\textbf {\bibinfo {volume} {90}},\ \bibinfo {pages}
  {085146} (\bibinfo {year} {2014})}\BibitemShut {NoStop}%
\bibitem [{\citenamefont {White}(1992)}]{white1992density}%
  \BibitemOpen
  \bibfield  {author} {\bibinfo {author} {\bibfnamefont {S.~R.}\ \bibnamefont
  {White}},\ }\href@noop {} {\bibfield  {journal} {\bibinfo  {journal}
  {Physical review letters}\ }\textbf {\bibinfo {volume} {69}},\ \bibinfo
  {pages} {2863} (\bibinfo {year} {1992})}\BibitemShut {NoStop}%
\bibitem [{\citenamefont {Peng}\ \emph {et~al.}(2023)\citenamefont {Peng},
  \citenamefont {Sheng},\ and\ \citenamefont
  {Jiang}}]{peng2023superconductivity}%
  \BibitemOpen
  \bibfield  {author} {\bibinfo {author} {\bibfnamefont {C.}~\bibnamefont
  {Peng}}, \bibinfo {author} {\bibfnamefont {D.}~\bibnamefont {Sheng}}, \ and\
  \bibinfo {author} {\bibfnamefont {H.-C.}\ \bibnamefont {Jiang}},\ }\href@noop
  {} {\bibfield  {journal} {\bibinfo  {journal} {arXiv preprint
  arXiv:2303.12348}\ } (\bibinfo {year} {2023})}\BibitemShut {NoStop}%
\bibitem [{\citenamefont {Zhao}\ \emph {et~al.}(2024)\citenamefont {Zhao},
  \citenamefont {Kang}, \citenamefont {Zhang}, \citenamefont {Kn{\"u}ppel},
  \citenamefont {Tao}, \citenamefont {Li}, \citenamefont {Tschirhart},
  \citenamefont {Redekop}, \citenamefont {Watanabe}, \citenamefont {Taniguchi}
  \emph {et~al.}}]{zhao2024realization}%
  \BibitemOpen
  \bibfield  {author} {\bibinfo {author} {\bibfnamefont {W.}~\bibnamefont
  {Zhao}}, \bibinfo {author} {\bibfnamefont {K.}~\bibnamefont {Kang}}, \bibinfo
  {author} {\bibfnamefont {Y.}~\bibnamefont {Zhang}}, \bibinfo {author}
  {\bibfnamefont {P.}~\bibnamefont {Kn{\"u}ppel}}, \bibinfo {author}
  {\bibfnamefont {Z.}~\bibnamefont {Tao}}, \bibinfo {author} {\bibfnamefont
  {L.}~\bibnamefont {Li}}, \bibinfo {author} {\bibfnamefont {C.~L.}\
  \bibnamefont {Tschirhart}}, \bibinfo {author} {\bibfnamefont
  {E.}~\bibnamefont {Redekop}}, \bibinfo {author} {\bibfnamefont
  {K.}~\bibnamefont {Watanabe}}, \bibinfo {author} {\bibfnamefont
  {T.}~\bibnamefont {Taniguchi}},  \emph {et~al.},\ }\href@noop {} {\bibfield
  {journal} {\bibinfo  {journal} {Nature Physics}\ ,\ \bibinfo {pages} {1}}
  (\bibinfo {year} {2024})}\BibitemShut {NoStop}%
\end{thebibliography}%

\newpage

\widetext
\begin{center}
\textbf{\large Supplemental Materials: Phase Diagram of Kane-Mele Hubbard model at small doping}
\end{center}
\setcounter{equation}{0}
\setcounter{figure}{0}
\setcounter{table}{0}
\setcounter{page}{1}
\makeatletter
\renewcommand{\theequation}{S\arabic{equation}}
\renewcommand{\thefigure}{S\arabic{figure}}
\renewcommand{\bibnumfmt}[1]{[S#1]}
\renewcommand{\citenumfont}[1]{S#1}

In this supplementary we provide more results to support our finding. We also provide analytical results to explain our findings. In Sec. A, we provide the details of the DMRG calculations. In section A of the supplementary, we discuss about the DMRG method used and more results to support out finding. in Section B, we show that RPA calculation also support out findings.

\section{Section A: Details of DMRG calculation}
\begin{figure}[H]
	\centering
        \includesvg[width=12cm]{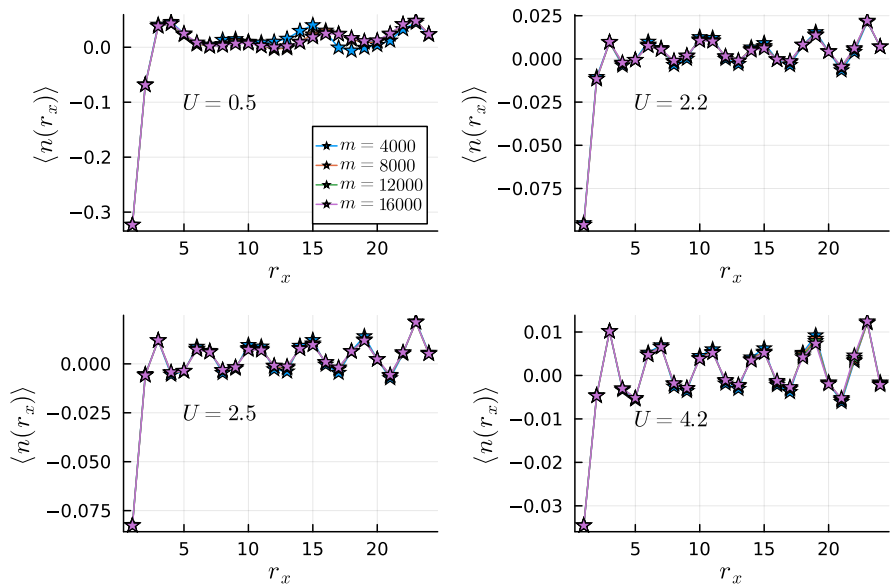}
	\caption{Charge density fluctuation along $x-$axis of the cylinder for different values of interaction strengths $U$ ranging from $0.5$ to $4.2$,   $\lambda= 0.2$ and different values of the bond dimensions ($m = 4000$ to $m=16000$). Average density is subtracted from the local density to emphasise on the small fluctuation which will go away with the system size.}
	\label{fig:supp_fig_1}
\end{figure}

Most of the calculations presented in the main text are obtained for a system size $ N = 24 \times 6$, and we kept bond dimension upto $ M = 16000$. We preformed DMRG starting with bond dimension $m=4000$ and gradually went upto $m=16000$. We performed $1/m$ scaling on most of the results presented in the main text to obtain $m=\infty$ limit.  In Fig. \ref{fig:supp_fig_1}, we show the charge density fluctuation along the $x-$axis of the cylinder (long axis of the cylinder) for $\lambda$=0.2 and different values of interaction strength $U$ and bond dimensions $m$. Average charge density is subtracted from the result to make the fluctuation easy to read. We find that the charge density fluctuation is relatively small. Interestingly, it is minimum at the center of the cylinder which indicates that it is a finite size effect and should go away in the thermodynamic limit.

\begin{figure}[H]
	\centering
        \includesvg[width=12cm]{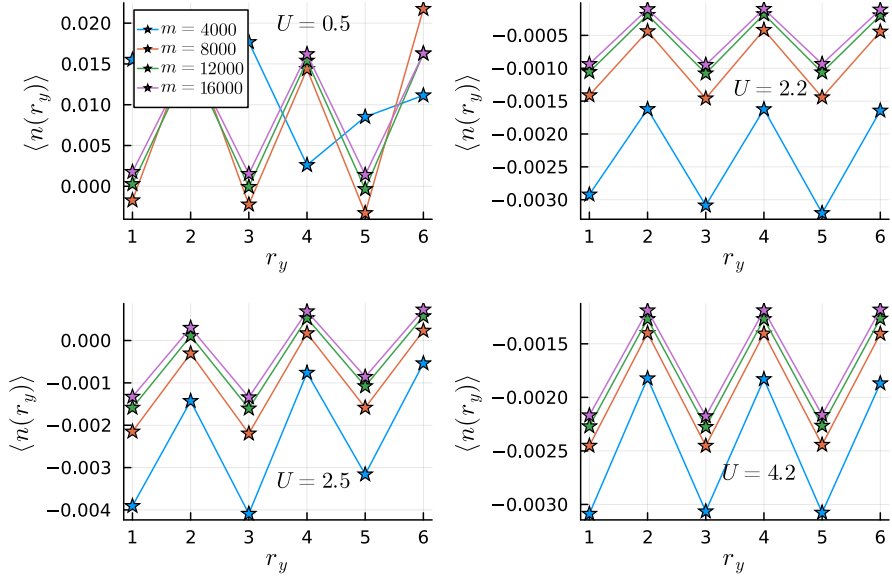}
	\caption{Charge density fluctuation along $y-$axis of the cylinder for different values of interaction strengths $U$ ranging from $0.5$ to $4.2$,   $\lambda= 0.2$ and different values of the bond dimensions ($m = 4000$ to $m=16000$). Average density is subtracted from the local density to emphasise on the small fluctuation which will go away with the system size.}
	\label{fig:supp_fig_3}
\end{figure}

In Fig. \ref{fig:supp_fig_3}, we show the similar charge density fluctuation as in Fig. \ref{fig:supp_fig_1} but along the $y-$axis of the cylinder(circumference of the cylinder) for different values of the interaction strength and bond dimension. Average of the total charge has been subtracted for better readability. Again, we find very small fluctuation in the charge which we think is due to finite size effect and should go away with increase in the system size and bond dimension.

\section{Section B: RPA calculation}
To perform the RPA calculation we start with the non-interacting Hamiltonian given in Eq. 1 of the main text. The wave-vector dependent susceptibility is given by
\begin{eqnarray}
    \chi_0^{\alpha,\beta}(q) = \sum_{k} \frac{f(\epsilon_{k+q}^\alpha) - f(\epsilon_k^\beta)}{(\epsilon_{k+q}^\alpha - \epsilon_k^\beta)}
\end{eqnarray}
Where $\epsilon_k^\alpha$ are the bands of the the Hamiltonian in Eq. 1 of the main text, $f(\epsilon_k^\alpha)$ is the Fermi function. We find that that $\chi_0^{\alpha,\beta}(q)$ in Eq.has strong peaks for some values of $q$ which increases as we increase the value of spin orbit coupling $\lambda$.  

In the Hubbard model, the pairing interaction mediated by paramagnon is obtained in the single and triplet channels by[]
\begin{eqnarray}
    V_s = \frac{U^2\chi_0(p^\prime+p)}{1-U\chi_0(p^\prime +p)} + \frac{U^3\chi^2_0(p^\prime-p)}{1-U^2\chi_0^2(p^\prime - p)}
\end{eqnarray}

and 

\begin{eqnarray}
    V_t = -\frac{U^2\chi_0(p^\prime-p)}{1-U^2\chi_0^2(p^\prime - p)}
\end{eqnarray}
Hence for the RPA susceptibility in the singlet channel to diverge, we need to satisfy 
\begin{eqnarray}
    1-U_c\chi_0(p) = 0\nonumber\\
    \implies U_c = \frac{1}{\chi_0(p)}
    \text{or}\nonumber\\
    1-U_c^2\chi_0^2(p) = 0\nonumber\\
    \implies U_c=\pm \frac{1}{\chi_0(q)}
\end{eqnarray}
as the value of $\chi_0(q)$ increases with $\lambda$, $U_c$ required for RPA susceptibility to diverge reduces and hence the critical interaction for superconducting transition also decreases.



\end{document}